\documentclass[english]{article}
\usepackage[latin9]{inputenc}
\usepackage{float}
\usepackage{amssymb}

\makeatletter

\providecommand{\tabularnewline}{\\}

\providecommand{\tabularnewline}{\\}
\usepackage{babel}

\makeatother

\usepackage{babel}
\begin{document}

\title{Schematic interactions with many degeneracies}

\author{Arun Kingan, Michael Quinonez and Larry Zamick\\
 Department of Physics and Astronomy\\
 Rutgers University, Piscataway, New Jersey 08854 }
\maketitle
\begin{abstract}
In previous works we examined the spectra for systems of 2 protons
and 2 neutrons, in a single j shell calculation, by obtaining matrix
elements from experiment. More recently we considered schematic interactions
in the same model space. We continue in this vein here. The present
work and the former can be regarded as 2 bookends on a bookshelf. 
\\
\\
PACs Number: 21.60.Cs
\end{abstract}

\section{Introduction}

In 1963 and 1964, calculations were performed to obtain wave functions
and energy levels in the f$_{7/2}$ shell by Bayman et al. {[}1{]},
McCullen et al. {[}2{]} and Ginocchio and French {[}3{]}. At that
time, the T=1 two-body matrix elements were well known but not so
the T=0. In 1985, the T=0 matrix elements were better known and the
calculations were repeated by Escuderos et al. {[}4{]}. The two-particle
matrix elements, obtained mainly from the spectrum of $^{42}$Sc,
are shown in Ref {[}4{]} . Note not only is J=0 T=1 low lying,
but J=1 T=0 and J=7 T=0.

We then considered {[}5{]} a schematic interaction E(J)=0 for odd
J (T=0) and E(J)= J for even J (T=1). For convenience, we here change
from J to J/2 as was done in ref. {[}6{]}. We also note work by K.
Neerg$\ddot{a}$rd ref {[}7{]}, which was important in getting analytical
results in ref {[}5{]}. We call this the 0123 interaction. In more
detail this can be defined by the input SET1 \{0,0,1,0,2,0,3,0\} Our
initial motivation was to find a 2-particle interaction which yielded
an equally spaced spectrum for even I states of a system of 2 protons
and 2 neutrons e.g. $^{44}$Ti. We were partly successful--the lowest
few states were not equally spaced, but then there was a critical
angular momentum beyond which we found equal spacing.

In this work we will consider another extreme. Rather than an equally
spaced spectrum for high angular momentum we get a collapse with many
degenerate levels. These 2 models will then be like the bookends between
which there will be less extreme cases.

As a physical motivation we note that this work puts greater emphasis
on the top of the spectrum rather than the more studied at the beginning.
As an interesting example compare the I=10$^{+}$ and 12$^{+}$ states
in $^{44}$Ti and $^{52}$Fe. In the single j shell model (f$_{7/2}$)
with a fixed interaction, the spectra of the 2 nuclei should be identical.
In $^{44}$Ti the 12$^{+}$ state is at 8039.9 keV and the 10$^{+}$
state is at 7671.1 keV, i.e. 368 keV below the 12$^{+}$ state. In
$^{52}$Fe the 10$^{+}$ state is at 7381.9 keV whilst the 12$^{+}$
state is 423.9 keV below at 6958.8 keV. Thus the 12$^{+}$ state cannot
decay by quadrupole radiation and is strongly isomeric. in $^{44}$Ti
the half-life of the 12$^{+}$ state is 2.1 ns (weakly isomeric) whilst
for $^{52}$Fe the corresponding value is 45.9 s. All this is discussed
in Ref. {[}8{]}.

In this work we get an exaggeration of this behavior, but hopefully
we gain some insight into the workings of complex shell model results.

\section{A new schematic interaction }

As a counterpoint to the 0123 interaction we consider here a new schematic
interaction 0122, or in more detail SET2 \{0,0,1,0,2,0,2,0{]}. That
is, we have E(J) = 0 for odd J whilst for even J E(J) is still equal
to J/2 for J $\leqq$ (2j-3) but E(2j-1) = E(2j-3). In the g$_{9/2}$
shell we have the 01233 interaction and in h$_{11/2}$ we have the
012344 interaction. We will later also consider the interaction \{0,0,1,0,2,0,C,0).
By choosing C=3 we retrieve the results of ref {[}4{]}. With C=2 we
get the new results and with C in between we see how the changes evolve.

\section{Spectra (MeV) of 2 protons and 2 neutrons in the f$_{7/2}$shell
--$^{44}$Ti.}

We show in Table I the spectra of 2 protons and 2 neutrons in the
f$_{7/2}$ shell. First we have the MBZE interaction, in which the
2 body matrix elements were taken from experiment, more specifically
from the spectrum of $^{42}$Sc. Then we show results for the previously
considered 0123 interaction and then new results for the 0122 interaction.

\begin{table}[H]
\centering \caption{Even I spectra in $^{44}$Ti for various interactions.}
\begin{tabular}{c c c c}
\hline 
I  & MBZE  & 0123  & 0122 \tabularnewline
\hline 
0  & 0.0000  & 0.0000  & 0.0000 \tabularnewline
2  & 1.1631  & 0.7552  & 0.8242\tabularnewline
4  & 2.7900  & 1.8330  & 1.9051\tabularnewline
6  & 4.0618  & 3.1498  & 2.1497 \tabularnewline
8  & 6.0842  & 4.6498  & 3.5221 \tabularnewline
10  & 7.3839  & 6.1498  & 4.7840 \tabularnewline
12  & 7.7022  & 7.6498  & 4.7840 \tabularnewline
\hline 
\end{tabular}
\end{table}

Note that with the MBZE interaction, the gap from I=12 to I =10 is
much smaller than from I=10 to I=8. With the 0123 interaction, however,
from I=6 on we get equally spaced spectra with 1.5 MeV gaps. That
is to say The 12-10 splitting is the same as the 10-8 splitting is
the same as the 8-6 splitting. This was discussed extensively in Ref
{[}5{]}.

We now consider the new result in the last column. In contrast to
what happens with the 0123 interactions, we find with 0122 there is
a collapse at the top, with I=12$^{+}$ and I = 10$^{+}$ degenerate.
Thus, the 2 interactions form the extremes, or two \char`\"{}bookends\char`\"{}
with equal spacing at one end and collapse at the other.

We note that the excitation energy 4.7840 MeV occurs many times, and
sometimes we have degenerate doublets.

To show what is happening we list all the wave functions and energies
of I=8$^{+}$, I=5$^{+}$, and I=10$^{+}$ states for 2 protons and
2 neutrons in the f$_{7/2}$ shell with the 0122 interaction in Table
II, IV, and V respectively; and in Table III we show a special state
of I=4$^{+}$.

\begin{table}[H]
\global\long\def\thetable{II}
 \centering \caption{Wave functions and energies I=8$^{+}$}
\setlength{\tabcolsep}{3pt} %
\begin{tabular}{c c c c c c c c}
\hline 
J$_{p}$  & J$_{n}$  & E=3.5221  & E=4.7840  & E=4.7840  & E=5.6691  & E=6.4942  & E=10.3078 \tabularnewline
\hline 
2.0  & 6.0  & 0.6486  & 0.0000  & -0.0000  & -0.6927  & 0.1421  & -0.2817 \tabularnewline
4.0  & 4.0  & 0.1652  & 0.9087  & 0.0488  & 0.0000  & 0.0000  & 0.3803 \tabularnewline
4.0  & 6.0  & 0.1594  & -0.2117  & 0.5433  & -0.1421  & -0.6927  & 0.3669 \tabularnewline
6.0  & 2.0  & 0.6486  & -0.0000  & -0.0000  & 0.6927  & -0.1421  & -0.2817 \tabularnewline
6.0  & 4.0  & 0.1594  & -0.2117  & 0.5433  & 0.1421  & 0.6927  & 0.3669 \tabularnewline
6.0  & 6.0  & -0.2840  & 0.2910  & 0.6381  & 0.0000  & -0.0000  & -0.6538 \tabularnewline
\hline 
\end{tabular}
\end{table}

\begin{table}[H]
\global\long\def\thetable{III}
 \centering \caption{I=4$^{+}$ Special State}
\begin{tabular}{c c c}
\hline 
J$_{p}$  & J$_{n}$  & E=4.7840\tabularnewline
\hline 
0.0  & 4.0  & 0.0000 \tabularnewline
2.0  & 2.0  & 0.0000 \tabularnewline
2.0  & 4.0  & 0.0000 \tabularnewline
2.0  & 6.0  & 0.0000 \tabularnewline
4.0  & 0.0  & 0.0000 \tabularnewline
4.0  & 2.0  & 0.0000 \tabularnewline
4.0  & 4.0  & -0.8870 \tabularnewline
4.0  & 6.0  & -0.1735 \tabularnewline
6.0  & 2.0  & 0.0000 \tabularnewline
6.0  & 4.0  & -0.1735 \tabularnewline
6.0  & 6.0  & 0.3912 \tabularnewline
\hline 
\end{tabular}
\end{table}

\begin{table}[H]
\global\long\def\thetable{IV}
 \centering \caption{Wave functions and energies I=5$^{+}$}
\setlength{\tabcolsep}{2pt} %
\begin{tabular}{c c c c c c}
\hline 
J$_{p}$  & J$_{n}$  & E=3.2840  & E=3.8554  & E=4.4950  & E=4.7528\tabularnewline
\hline 
2.0  & 4.0  & -0.4707  & 0.4151  & -0.3356  & 0.4850 \tabularnewline
2.0  & 6.0  & 0.5276  & 0.3704  & 0.5492  & 0.3708 \tabularnewline
4.0  & 2.0  & 0.4707  & -0.4151  & -0.3356  & 0.4850 \tabularnewline
4.0  & 4.0  & -0.0000  & 0.0000  & 0.1239  & 0.0902 \tabularnewline
4.0  & 6.0  & -0.0000  & 0.4364  & 0.1922  & 0.1098 \tabularnewline
6.0  & 2.0  & -0.5276  & -0.3704  & 0.5492  & 0.3708 \tabularnewline
6.0  & 4.0  & 0.0000  & -0.4364  & 0.1922  & 0.1098\tabularnewline
6.0  & 6.0  & 0.0000  & 0.0000  & -0.2868  & 0.4714 \tabularnewline
\hline 
  &  &  &  &  & \tabularnewline
\hline 
J$_{p}$  & J$_{n}$  & E=5.8703  & E=6.1990  & E=6.6387  & E=9.6411 \tabularnewline
\hline 
2.0  & 4.0  & -0.3513  & -0.1334  & -0.1047  & -0.3257 \tabularnewline
2.0  & 6.0  & 0.0966  & -0.1878  & -0.1274  & -0.2905 \tabularnewline
4.0  & 2.0  & -0.3513  & -0.1334  & -0.1047  & 0.3257 \tabularnewline
4.0  & 4.0  & -0.1488  & 0.8221  & -0.5278  & 0.0000 \tabularnewline
4.0  & 6.0  & -0.2927  & 0.2572  & 0.5470  & 0.5563 \tabularnewline
6.0  & 2.0  & 0.0966  & -0.1878  & -0.1274  & 0.2905\tabularnewline
6.0  & 4.0  & -0.2927  & 0.2572  & 0.5470  & -0.5563 \tabularnewline
6.0  & 6.0  & 0.7356  & 0.2929  & 0.2620  & 0.0000 \tabularnewline
\hline 
\end{tabular}
\end{table}

\begin{table}[H]
\global\long\def\thetable{V}
 \centering \caption{Wave functions and energies I=10$^{+}$}
\setlength{\tabcolsep}{3pt} %
\begin{tabular}{c c c c c}
\hline 
Jp  & Jn  & E=4.7840  & E=4.7840  & E=6.7840 \tabularnewline
\hline 
4.0  & 6.0  & 0.6704  & -0.2249  & 0.7071 \tabularnewline
6.0  & 4.0  & 0.6704  & -0.2249  & -0.7071 \tabularnewline
6.0  & 6.0  & 0.3180  & 0.9481  & -0.0000 \tabularnewline
\hline 
\end{tabular}
\end{table}

Note that the amplitude D(J$_{p}$ J$_{n}$) is either plus or minus
D(J$_{n}$ J$_{p}$). This is due to charge symmetry of the 2-body
interaction. In general, we have for the N=Z nucleus:

\centerline{D$^{IT}$(J$_{p}$J$_{n}$)= (-1) $^{(I+T)}$D$^{IT}$(J$_{n}$J$_{p}$)}
\setlength{\parindent}{0cm} where I is the total angular momentum
and T is the isospin. \setlength{\parindent}{0.5cm}

Clearly then the 5.6091 and 6.4942 MeV states are T=1 states and we
will defer discussing them until later. Less obvious is that the 10.3078
state has isospin T=2. It is the double analog of the unique T=2 J=8$^{+}$
state in $^{44}$Ca.

\section{Degeneracies and 2 particle fractional parentage coefficients}

We now discuss this in a more systematic way. There are many degeneracies
in the new interaction, they are listed in the tables IX, X, and XI
for the f7/2, g9/2, and h11/2 shells respectively. We can correlate
these with the number of T=2 states in the last columns of tables
VI, VII and VIII. These can easily be obtained from the work of Bayman
and Lande ref {[}9{]}.

\begin{table}[H]
\global\long\def\thetable{VI}
 \centering \caption{Occurrence of special energies: f$_{7/2}$}
\setlength{\tabcolsep}{3pt} %
\begin{tabular}{c c c c c}
\hline 
I  & 3.2840  & 4.7840  & 6.7840  & \# of T=2  \tabularnewline
\hline 
0  &  & 1  &  & 1 \tabularnewline
1  &  &  &  &  \tabularnewline
2  &  & 1  &  & 2 \tabularnewline
3  & 1  & 1  &  &  \tabularnewline
4  &  & 1  &  & 2  \tabularnewline
5  & 1  &  &  & 1 \tabularnewline
6  & 1  & 2  &  & 1 \tabularnewline
7  & 1  & 1  & 1  &  \tabularnewline
8  &  & 2  &  & 1 \tabularnewline
9  &  & 1  & 1  & \tabularnewline
10  &  & 2  & 1  & \tabularnewline
11  &  &  & 1  & \tabularnewline
12  &  & 1  &  & \tabularnewline
\hline 
\end{tabular}
\end{table}

\begin{table}[H]
\global\long\def\thetable{VII}
 \centering \caption{Occurrence of Special Energies g$_{9/2}$}
\setlength{\tabcolsep}{3pt} %
\begin{tabular}{c c c c c c}
\hline 
I  & 4.3644  & 5.8644  & 7.3644  & 10.3644  & \# of T=2 \tabularnewline
\hline 
0  &  &  &  &  & 2  \tabularnewline
1  &  &  &  &  &  \tabularnewline
2  &  &  & 1  &  &   \tabularnewline
3  &  &  &  &  & 1  \tabularnewline
4  &  &  &  &  & 3  \tabularnewline
5  &  & 1  &  &  & 1 \tabularnewline
6  &  &  &  &  & 3  \tabularnewline
7  & 1  & 1  &  &  & 1 \tabularnewline
8  & 1  &  & 1  &  & 2  \tabularnewline
9  &  & 1  &  &  & 1 \tabularnewline
10  &  & 1  & 2  &  & 1 \tabularnewline
11  &  & 1  & 1  & 1  & \tabularnewline
12  &  &  & 2  &  & 1 \tabularnewline
13  &  &  & 1  & 1  & \tabularnewline
14  &  &  & 2  & 1  &  \tabularnewline
15  &  &  &  & 1  & \tabularnewline
16  &  &  & 1  &  &  \tabularnewline
\hline 
\end{tabular}
\end{table}

\begin{table}[H]
\global\long\def\thetable{VIII}
 \centering \caption{Occurrence of Special Energies h$_{11/2}$}
\setlength{\tabcolsep}{3pt} %
\begin{tabular}{c c c c c c c}
\hline 
I  & 5.5137  & 7.0137  & 8.5137  & 10.0137  & 14.0137  & \# of T=2  \tabularnewline
\hline 
0  &  &  &  &  &  & 2 \tabularnewline
1  &  &  &  &  &  &  \tabularnewline
2  &  &  &  &  &  & 3 \tabularnewline
3  &  &  &  &  &  & 1  \tabularnewline
4  &  &  &  &  &  & 4  \tabularnewline
5  &  &  &  &  &  & 2  \tabularnewline
6  &  &  &  &  &  & 4  \tabularnewline
7  &  &  &  &  &  & 2  \tabularnewline
8  &  &  &  &  &  & 4  \tabularnewline
9  & 1  &  &  &  &  & 2 \tabularnewline
10  &  &  &  &  &  & 3 \tabularnewline
11  &  & 1  & 1  &  &  & 1 \tabularnewline
12  &  & 1  &  & 1  &  & 2 \tabularnewline
13  &  &  & 1  &  &  & 1 \tabularnewline
14  &  &  & 1  & 2  &  & 1 \tabularnewline
15  &  &  & 1  & 1  & 1  & \tabularnewline
16  &  &  &  & 2  &  & 1 \tabularnewline
17  &  &  &  & 1  & 1  & \tabularnewline
18  &  &  &  & 2  & 1  &  \tabularnewline
19  &  &  &  &  & 1  &  \tabularnewline
20  &  &  &  & 1  &  &  \tabularnewline
\hline 
\end{tabular}
\end{table}

Let us now focus on the 2-fold degenerate doublet at 4.7840 MeV for
I=8$^{+}$. We know that the non-vanishing amplitudes involve angular
momenta 4 and 6. There are 4 non-vanishing amplitudes, but only 3
are independent D(44), D(46) (which is equal to D(64)) and D(66).Why
this strange behavior? Since for the 0122 interaction E(4)=E(6) we
see that the interaction is effectively a constant in the limited
J=4 and 6 state. But there is a constraint: for these 2 states to
be T=0 states, they must be orthogonal to the single T=2 state. There
is one more constraint: normalization.\\
 \centerline{D(44)$^{2}$ +2 D(46)$^{2}$+ D(66)$^{2}$= 1.}\\
 It is easy to see that there are 2 solutions since we have one more
parameter than there are constraints. Thus we have a degenerate doublet.

In contrast for I=4$^{+}$ we have only one special state. This is
because there are 2 isospin T=2 states for I=4$^{+}$, thus 3 parameters
and 3 constraints: only one solution.

Let us now consider odd I states and focus on I=5$^{+}$. We now have
a quite different behavior. There is only one special state and it
is at 3.2840 MeV, 1.5 MeV lower than the special I=8$^{+}$states.
This is because for odd I, T=0 states, D(44)=D(66)=0 so there is only
one parameter D(46) to play with. We are unable to construct a state
orthogonal to the lone T=2 state with these limited configurations.
However, with the configurations 24 and 26 we have 2 independent parameters
and 2 constraints: normalization and orthogonality to the lone I=5$^{+}$
T=2 state. Thus, there is only one possible solution.

Note that all the "special sates" with T=0 all have the same energy: 4.784 MeV.
Things are perhaps clearer if we make the lowest I=0, T=0 energy to be -4.784 MeV.
Then all the special states have zero energy. We would get this result with
a Hamiltonian H=0, and this is the Hamiltonian that this subclass of states sees.
Note that all the T=0 special states are linear combinations of basis states with (Jp,Jn)
(4,4), (4,6+6,4) and (6,6). The coefficients have been obtained by making them
orthogonal to T=2 states. It can be shown that they are eigenstates of the Hamiltonian
by noting one cannot add a component like say (2,6+6,2) to the special wave function.
This new state would not be orthogonal to a T=2 state and hence would be a mixture
of T=0 and T=2. Some more details can be found in Rule 2 of the previous publication [5].

\section{From old to new}

In previously published work {[}5,6{]} with the 0123, 01234, and 012345
interactions for the f$_{7/2}$, g$_{9/2}$ and h$_{11/2}$ shells
respectively, we find critical angular momentum beyond which we get
equally spaced spectra. The values are respectively 6, 8, and 10 with
an obvious generalization to higher shells.

\begin{table}[H]
\global\long\def\thetable{IX}
 \centering \caption{Special states in the f$_{7/2}$ shell (0123 interaction)}
\setlength{\tabcolsep}{3pt} %
\begin{tabular}{c c c}
\hline 
J$_{p}$ + J$_{n}$  & E (MeV)  & I  \tabularnewline
\hline 
6  & 3.15  & 3, 6 \tabularnewline
8  & 4.65  & 6, 7, 8 \tabularnewline
10  & 6.15  & 3, 7, 9, 10 \tabularnewline
12  & 7.65  & 10, 12 \tabularnewline
\hline 
\end{tabular}
\end{table}

\begin{table}[H]
\global\long\def\thetable{X}
 \centering \caption{Special states in the g$_{9/2}$ shell (01234 interaction)}
\setlength{\tabcolsep}{3pt} %
\begin{tabular}{c c c}
\hline 
J$_{p}$ + J$_{n}$  & E (MeV)  & I \tabularnewline
\hline 
8  & 4.29  & 8 \tabularnewline
10  & 5.79  & 7, 9,10 \tabularnewline
12  & 7.29  & 10, 11, 12 \tabularnewline
14  & 8.79  & 11, 13, 14 \tabularnewline
16  & 10.29  & 14, 16 \tabularnewline
\hline 
\end{tabular}
\end{table}

\begin{table}[H]
\global\long\def\thetable{XI}
 \centering \caption{Special states in the h$_{11/2}$ shell (012345 interaction))}
\setlength{\tabcolsep}{3pt} %
\begin{tabular}{c c c}
\hline 
J$_{p}$ + J$_{n}$  & E (MeV)  & I \tabularnewline
\hline 
10  & 5.46  & 10 \tabularnewline
12  & 6.96  & 11, 12 \tabularnewline
14  & 8.46  & 11, 13, 14 \tabularnewline
16  & 9.96  & 14, 15, 16 \tabularnewline
18  & 11.46  & 15, 17, 18 \tabularnewline
20  & 12.96  & 18, 20 \tabularnewline
\hline 
\end{tabular}
\end{table}

Tables IX, X and XI show the equally spaced spectra (1.5 MeV gaps)
in the f$_{7/2}$, g$_{9/2}$ and h$_{11/2}$ shells respectively,
as well as the angular momenta that belong to these states.

Let us focus on the I=12$^{+}$--I=10$^{+}$ splitting in the f$_{7/2}$
shell. With the old interaction, SET1, this splitting is 1.5 MeV while
with the current interaction, SET2, the states are degenerate. With
an interaction SET3=\{0,0,1,0,2,0,C,0\} we find the splitting is 1.5C.
That is to say, the splitting is linear in C. This behavior is also found 
in higher shells. Specifically in the g$_{9/2}$ shell with the interaction
\{0,0,1,0,2,0,3,0,C,0\}, the J=16 and J=14 splitting is also proportional to C.

Some of the results can be explained by work in of Robinson and Zamick
{[}8,9{]}. This pertains to states for which (J$_{p}$,J$_{n}$) are
good quantum numbers, i.e. T=0 states with angular momenta which do
not occur for T=2 states. In the f$_{7/2}$ shell, these angular momenta
are 3, 7, 9, 10, and 12. Let us look at I=10$^{+}$. The absence of
the coupling between (4,6) and (6,6) in both SET1 and SET2 was explained
in the early work {[}8.10,11{]} and is shown by the vanishing of the unitary
9j coefficient ((7/2,7/2)$^{6}$ (7/2,7/2)$^{6}$\textbar{} (7/2,7/2)$^{6}$(7/2
,7/2)$^{4}$)$^{10}$.

The special states of SET1 go beyond this and are characterized by
having wavefunctions such that (J$_{p}$+J$_{n}$) is a constant.
This is discussed in ref{[}5{]}.

\section{Level Inversions}

Although it is somewhat out of the scope of the model we are discussing here,
an interesting phenomenon is level inversion at the high end of the spectrum.
One of us has previously discussed this in [7] and [14] so our discussion here
will be brief. For example in $^{44}$Ti, the J=10+ state is lower in energy than J=12+
but they are sufficiently close so that the J=12+ state is isomeric. However, in
 $^{52}$Fe there is an inversion with J=12+ below J=10+. Since no B(E2) is possible, this
12+ state is strongly isomeric.

In the single j shell model with the same interaction these 2 nuclei would have 
identical spectra since one has 2 protons and 2 neutrons and the other the proton 
holes and 2 neutron holes. To get changes one has to use different interactions. In 
ref [14] table 7 we see that for $^{44}$Ti the J=J$_{max}$=7 2 body matrix element is 0.6163 
MeV above the J=0 2 body matrix element, whereas for $^{52}$Fe it is 0.1999 MeV. 
Lowering J=J$_{max}$ helps to create level inversion. Indeed, investigating the spectrum 
of the 2 hole system $^{54}$Co, one sees that this splitting is smaller than in $^{44}$Ti.

Other nuclei are also considered, e.g. the $^{44}$Sc and $^{52}$Mn; and in the g9/2 shell 
 $^{96}$Cd and $^{96}$Ag

We now briefly consider how the splitting V(6)-V(4) for the 2 particle system affects the splitting
E(12)-E(10) for the 4 particle system. With the interaction INTa relevant for the $^{44}$Ti calculation
we have V(6)=3.242 MeV and V(4)=2.815 MeV, hence V(6)-V(4)=0.427 MeV. The splitting in
$^{44}$Ti is E(12)-E(10)=0.282 MeV. The corresponding numbers for the INTb interaction relevant to
$^{52}$Fe are 2.960 MeV, 2.645 MeV, 0.325 MeV and E(12)-E(10)= -0.122 MeV. If we now keep 
the V(J) in INTb as is, except we modify V(6) by making V(6)-V(4)=0.427. That is we assume 
the INTa gap. This makes V(6)=3.072 MeV. We now find E(12)-E(10)=+0.038 MeV.

Experimentally E(12)-E(10) is negative so we see that lowering the gap V(6)-V(4)
in going from INTa to INTb is important for obtaining the spin reversal.

\begin{table}[H]
\global\long\def\thetable{XII}
\centering \caption{E(12)-E(10) MeV Splitting }
\setlength{\tabcolsep}{3pt} %
\begin{tabular}{c c c c c}
\hline 
 & V(4) & V(6) & V(6)-V(4) & E(12)-E(10)\tabularnewline
\hline 
INTa($^{44}$Ti) & 2.815 & 3.242 & 0.427 & +0.282 \tabularnewline
INTb($^{52}$Fe) & 2.645 & 2.960 & 0.325 & -0.122 \tabularnewline
Mod & 2.645  & 3.072 & 0.427 & +0.038 \tabularnewline
\hline 
\end{tabular}
\end{table}

\section{Closing remarks}

There are schematic interactions with many more degenracies than what
we have found here. For example we haves the J=0 T=1 pairing interaction
of Flowers and Edmonds {[}12,13{]} which were recently used by one
of us (L.Z.) to explain the \char`\"{}gaps in nuclear spectra as traces
of seniority changes.\char`\"{} {[}15{]}. In {[}12.13{]} seniority
is a good quantum number (as well as isospin and reduced isospin).
The word \char`\"{}traces\char`\"{} in the title of {[}15{]} suggests
how to properly make use of schematic interactions. In the realisitc
case seniority is not a good quantum number but remnants of it have
some effects on the nuclear spectra. We hope the models that we have
presented here will also help to cast some insight into the behaviours
of complex nuclear spectra.

\setcounter{secnumdepth}{0}

\section{Acknowledgements}

A.K. received support from the Rutgers Aresty Summer 2016 program and the Richard J. Plano Research Award Summer 2017.
M.Q. has two institutional affiliations: Rutgers University and Central
Connecticut State University (CCSU). He received support via the Research
Experience for Undergraduates Summer program (REU) from the U.S. National
Science program Grant Number 1263280.

We are grateful to Kai Neerg$\ddot{a}$rd for his interest and help.

\end{document}